\documentclass[twocolumn,aps,10pt,showpacs,showkeys,]{revtex4}
\usepackage[dvips]{graphicx}

\begin{document}
\title{Electronic structure and crystal-field states in V$_{2}$O$_{3}$$^\spadesuit$}
\author{Z. Ropka}
\affiliation{Center of Solid State Physics, S$^{nt}$Filip 5, 31-150 Krakow, Poland,\\
Institute of Physics, Pedagogical University, 30-084 Krakow,
Poland}
\author{R. J. Radwanski}
\affiliation{Center of Solid State Physics, S$^{nt}$Filip 5,
31-150 Krakow, Poland\\Institute of Physics, Pedagogical
University, 30-084 Krakow, Poland}
\homepage{http://www.css-physics.edu.pl}
\email{sfradwan@cyf-kr.edu.pl}

\begin{abstract}
We have calculated the electronic structure of V$_2$O$_3$
associated with the V$^{3+}$ ions taking into account strong
on-site electron correlations and the spin-orbit coupling.
Closely lying 9 states of the subterm $^{3}$T$_{1g}$ are a
physical reason for exotic phenomena of V$_2$O$_3$. Electronic
structure and magnetism of V$^{3+}$ ions in the octahedral
surroundings are strongly susceptible to lattice distortions and
magnetic interactions. Our approach accounts both for the
insulating ground state, magnetism, including its orbital
contribution, as well as thermodynamical properties.

\pacs{71.70.-d, 75.10.Dq} \keywords{$V_2 O_3$, crystal field,
spin-orbit coupling, strong correlations}
\end{abstract}
\maketitle

 V$_2$O$_{3}$ attracts a great scientific interest by more than 50
years \cite{1}. Despite of it there is still strong discussion
about the description of its properties and its electronic
structure. A controversy for V$_2$O$_3$ starts already with the
electronic ground state. Within a localized paradigm there is a
long-standing controversy between a {\it S}=1 model without an
orbital degeneracy \cite{2} and the historically first $S$=1/2
orbitally degenerate model of Castellani et al.\cite{1}.  A spin-1
model with three degenerate orbitals was worked out by Di Matteo
et al. \cite{3}, whereas Refs \cite{4,5} develop a model, where
the fluctuations of t$_{2g}$ orbitals and frustrations play
prominent role in magnetism of a sister compound LaVO$_3$.

It is agreed that V$_2$O$_{3}$ is an insulating antiferromagnet
with the Neel temperature of 155-160 K. The basis of all theories
is the description of the V$^{3+}$ ion (2 d electrons) and its
electronic structure in the corundum structure, in which the V
ions are placed in a slightly distorted oxygen octahedron.

The aim of this paper is to present a low-energy electronic
structure of V$_2$O$_{3}$ as originating from the atomic-like
electronic structure of the strongly-correlated 3$d^{2}$
electronic systems occurring in the V$^{3+}$ ions. In our
description local atomic-scale effects, the orbital magnetism and
the intra-atomic spin-orbit coupling play the fundamentally
important role.
\begin{figure}[ht]
\begin{center}
\includegraphics[angle=0,width = 5.5 cm]{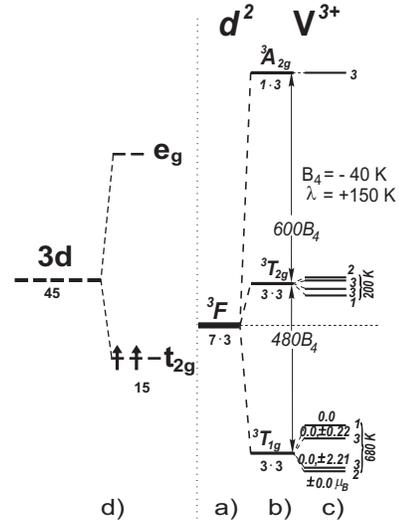}
\end{center} \vspace {-0.3 cm}
\caption{Lowest part of the electronic structure of V$_2$O$_3$
associated with the d$^2$ (V$^{3+}$) ion in the
strongly-correlated limit. a) Hund's rules ground term $^3F$, b)
splitting by the octahedral CEF, c) many-electron CEF states in
the octahedral CEF in the presence of the intra-atomic spin-orbit
coupling: B$_4$= -40$~$K (=10Dq =600B$_4$= 2.06 eV) and the
spin-orbit coupling as is observed in the free ion, i.e.
$\lambda_{s-o}$= +150 K. d) a schematic electronic structure
customarily recalled in the current literature - this
single-electron electronic structure is, according to us,
completely oversimplified.}
\end{figure}
According to our approach to 3d compounds, that seems to be very
natural, but is far from being accepted within the present
magnetic solid-state community, each V ion in V$_2$O$_{3}$ has
two $d$ electrons that form a strongly correlated atomic-like
3$d^{2}$ electron system. These strong correlations among the
3$d$ electrons we account for by two Hund's rules, that yield the
$^{3}F$ ground term, Fig. 1a. In the oxygen octahedron
surroundings, realized in the corundum structure of V$_2$O$_{3}$,
the $^{3}F$ term splits into two orbital triplets $^{3}$T$_{1g}$,
$^{3}$T$_{2g}$ and an orbital singlet $^{3}A_{2g}$. In the
octahedron of negative oxygen ions the ground subterm is the
orbital triplet $^{3}$T$_{1g}$ as is shown in Fig. 1b. All
orbital states are triply spin degenerated. This degeneration is
removed by the intra-atomic spin-orbit coupling that is always
present in the ion/atom. The effect of the spin-orbit coupling is
shown in Fig. 1c. For low-temperature properties the nine states
of the $^{3}$T$_{1g}$ subterm are important. Off-octahedral
distortions cause further splitting of the shown states -
important is that there is no more low-energy states as only
shown. The present calculations have been performed with a
realistic octahedral CEF parameter  The spin-orbit coupling
effect amounts only to 3$\%$ the CEF effect but it has enormous
influence on the eigen-functions and low-temperature properties.

The derived many-electron electronic structure of the d$^2$
configuration is completely different from very simplified
electronic structure, shown in Fig. 1d, usually presented in the
current literature with two parallel spin-electrons in the triply
degenerated t$_{2g}$ orbitals. We do not know, why such an
oversimplified scheme is usually shown though the many-electron
CEF approach is known already by almost 70 years (it is true that
it was rather used not to a solid, but to 3d ions as impurities
in Electron Paramagnetic Resonance (EPR) experiments) \cite{6,7}.
A $t^2_{2g}$ state, usually recalled in the literature, is
related to the very strong crystal-field limit notation and is
15-fold degenerated. It contains 4 subterms: $^3$T$_{1g}$,
$^1$T$_{2g}$, $^1$E$_g$, $^1$A$_{1g}$ also with the lowest
$^3$T$_{1g}$ subterm. In our approach to 3d-ion compounds we
assume the CEF to be stronger than the spin-orbit coupling, but
not so strong to destroy intra-atomic coulombic correlations.
Thus we assume an on-site intermediate crystal-field limit,
keeping the atomic identity of the involved cation, to be
physically adequate for description of 3d-ion compounds.

\begin{figure}[ht]
\begin{center}
\includegraphics[width = 5.5 cm]{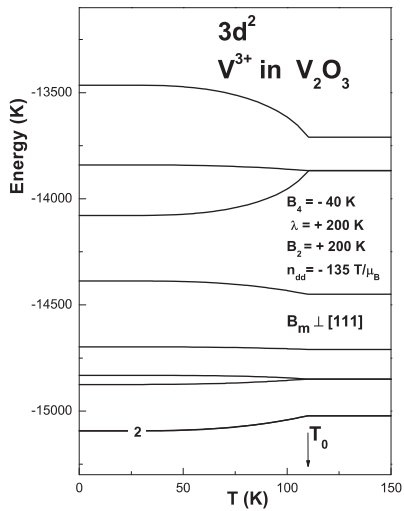}
\end{center} \vspace {-0.8 cm}
\caption{Modification of the electronic structure (9 lowest
states) of the d$^2$ (V$^{3+}$) ion in a magnetic state. The
parameters shown, with the trigonal distortion B$_2$= 200$~$K,
yield a magnetic moment of 1.19 $\mu_B$ close to the experimental
value of 1.2$\mu_B$ and the magnetic ordering temperature of 110
K. }
\end{figure}
For a magnetic state self-consistent calculations have been
performed similarly to those presented in Ref. \cite{8} for
FeBr$_{2}$. A value of 1.19 $\mu_ B $ is in perfect agreement
with experimental observation (1.2 $\mu_ B $) but the obtained
ordering temperature of 110 K is too low. Due to 5 closely lying
discrete states the electronic structure and magnetism are very
susceptible to lattice distortions and details of magnetic
interactions. These five closely lying states are a physical
reason for exotic experimental phenomena, when the V ion with such
complex local structure interacts in a solid.

In conclusion, we have calculated the electronic structure of
V$_2$O$_3$ associated with the V$^{3+}$ ions taking into account
strong on-site electron correlations and the spin-orbit coupling.
On basis of our studies we reject a possibility for an orbital
liquid in V$_2$O$_3$. V$^{3+}$ ions in the octahedral
surroundings are strongly susceptible to lattice distortions and
magnetic interactions. Our approach accounts both for the
insulating ground state, magnetism as well as thermodynamical
properties.

Answering a note of the SCES-05 first referee "While authors are
right, that the single electron theory cannot fully describe many
body effects, their schematic figure 1d is wrong. The point is,
that one does not speak about $E_{g}$ and $T_{2g}$ "levels", but
about $E_{g}$- and $T_{2g}$-like hybridizing "bands" of width few
eV." we maintain that we do not believe that so wide bands exist
in V$_{2}$O$_{3}$.

An extra remark. Despite the general critics of anonymous
referees and discriminating policy of Editors of Phys. Rev. B and
Phys. Rev. Lett., not mention others, we are convinced that such
strongly-correlated CEF-based approach is the proper starting
point for the physically adequate description of $3d$ compounds.
We call this approach Quantum Atomistic Solid State Theory
(QUASST) - for more details about QUASST (cond-mat/0010081) and
its application to a numerous compounds, e.g. LaCoO$_{3}$, CoO,
NiO, FeBr$_{2}$, LaMnO$_{3}$, one can find in our numerous papers
in ArXiv and Z. Ropka, and R. J. Radwanski, Phys. Rev. B {\bf67}
(2003) 172401. It is obvious to remind that the free exchange of
scientific information is essential for the development of Science.\\

$^\spadesuit$ Dedicated to the Pope John Paul II, a man of
freedom and truth in life and in Science.

\end{document}